\documentclass[preprint, 12pt,author-year,titlepage]{elsarticle} 
\usepackage[hyphens]{url}
\biboptions{sort&compress} 
\usepackage{graphicx}
\usepackage{booktabs} 
\makeatletter
\def\ps@pprintTitle{%
 \let\@oddhead\@empty
 \let\@evenhead\@empty
 \def\@oddfoot{\it \hfill\today}%
 \let\@evenfoot\@oddfoot}
\makeatother

\usepackage[T1]{fontenc}
\usepackage{lmodern}
\usepackage{amssymb,amsmath}
\usepackage{ifxetex,ifluatex}
\usepackage{fixltx2e} 
\IfFileExists{upquote.sty}{\usepackage{upquote}}{}
\ifnum 0\ifxetex 1\fi\ifluatex 1\fi=0 
  \usepackage[utf8]{inputenc}
\else 
  \usepackage{fontspec}
  \ifxetex
    \usepackage{xltxtra,xunicode}
  \fi
  \defaultfontfeatures{Mapping=tex-text,Scale=MatchLowercase}
  
\fi
\IfFileExists{microtype.sty}{\usepackage{microtype}}{}
\usepackage[verbose,letterpaper,tmargin=2.54cm,bmargin=2.54cm,l margin=2.54cm,rmargin=2.54cm]{geometry}
\usepackage{graphicx}
\makeatletter
\def\maxwidth{\ifdim\Gin@nat@width>\linewidth\linewidth
\else\Gin@nat@width\fi}
\makeatother
\let\Oldincludegraphics\includegraphics
\renewcommand{\includegraphics}[1]{\Oldincludegraphics[width=\maxwidth]{#1}}
\ifxetex
  \usepackage[setpagesize=false, 
              unicode=false, 
              xetex]{hyperref}
\else
  \usepackage[unicode=true]{hyperref}
\fi
\hypersetup{breaklinks=true,
            bookmarks=true,
            pdfauthor={},
            pdftitle={Optimal management of a stochastically varying population when policy adjustment is costly},
            colorlinks=true,
            urlcolor=blue,
            linkcolor=magenta,
            pdfborder={0 0 0}}
\begin{document}
\begin{frontmatter}

  \title{Optimal management of a stochastically varying population when policy
adjustment is costly}
    \author[UCB]{Carl Boettiger\corref{c1}}
   \ead{cboettig (at) berkeley.edu} 
   \cortext[c1]{Corresponding author}
    \author[melbourne]{Michael Bode}

    \author[UCD]{James N. Sanchirico}

    \author[UTKEcon]{Jacob LaRiviere}

    \author[UCD]{Alan Hastings}

    \author[UTKEEB]{Paul R. Armsworth}

      \address[UCB]{Department of Environmental Science, Policy and Management, University
of California, Berkeley, 130 Mulford Hall \#3114, Berkeley, CA
94720-3114, USA}    
    \address[UCD]{Department of Environmental Science and Policy, University of
California, Davis}    
    \address[melbourne]{School of Botany, University of Melbourne, Australia}    
    \address[UTKEEB]{Department of Ecology and Evolutionary Biology, University of Tennessee,
Knoxville}    
    \address[UTKEcon]{Department of Economics, University of Tennessee, Knoxville}    
  
  \begin{abstract}
  Ecological systems are dynamic and policies to manage them need to
  respond to that variation. However, policy adjustments will sometimes be
  costly, which means that fine-tuning a policy to track variability in
  the environment very tightly will only sometimes be worthwhile. We use a
  classic fisheries management question -- how to manage a stochastically
  varying population using annually varying quotas in order to maximize
  profit -- to examine how costs of policy adjustment change optimal
  management recommendations. Costs of policy adjustment (here changes in
  fishing quotas through time) could take different forms. For example,
  these costs may respond to the size of the change being implemented, or
  there could be a fixed cost any time a quota change is made. We show how
  different forms of policy costs have contrasting implications for
  optimal policies. Though it is frequently assumed that costs to
  adjusting policies will dampen variation in the policy, we show that
  certain cost structures can actually increase variation through time. We
  further show that failing to account for adjustment costs has a
  consistently worse economic impact than would assuming these costs are
  present when they are not.
  \end{abstract}
 \end{frontmatter}

\begin{flushleft}
\section{Introduction}\label{introduction}

Ecosystems are dynamic and exhibit rich patterns of variability in both
time and space (Durrett and Levin 1994). In designing management
policies for ecosystems, managers need to decide how much of that
variation to respond to. Managers could try to track variations in
ecosystem dynamics very closely by setting policies that are extremely
responsive to the environment. Many theoretical studies that seek to
identify optimal policies for exploited populations and communities
adopt this approach and largely ignore the challenges that would be
involved in implementing such recommendations (e.g., Reed 1979, Neubert
2003, Sethi et al. 2005, Halpern et al. 2011). However, the policy
process can often be much more sluggish to respond to variations in
ecosystem dynamics (Walters 1978, Armsworth et al. 2010). Moreover,
stakeholders impacted by ecosystem management may prefer some stability
and not want to deal with continually changing management
recommendations (Biais 1995, Armsworth and Roughgarden 2003, Patterson
2007, Patterson and Resimont 2007, Sanchirico et al. 2008). In other
words, whatever gains are available from fine-tuning a policy
prescription to more closely reflect environmental variation should be
traded off against potential costs associated with the more responsive
approach to management this would require.

\begin{figure}[htbp]
\centering
\includegraphics{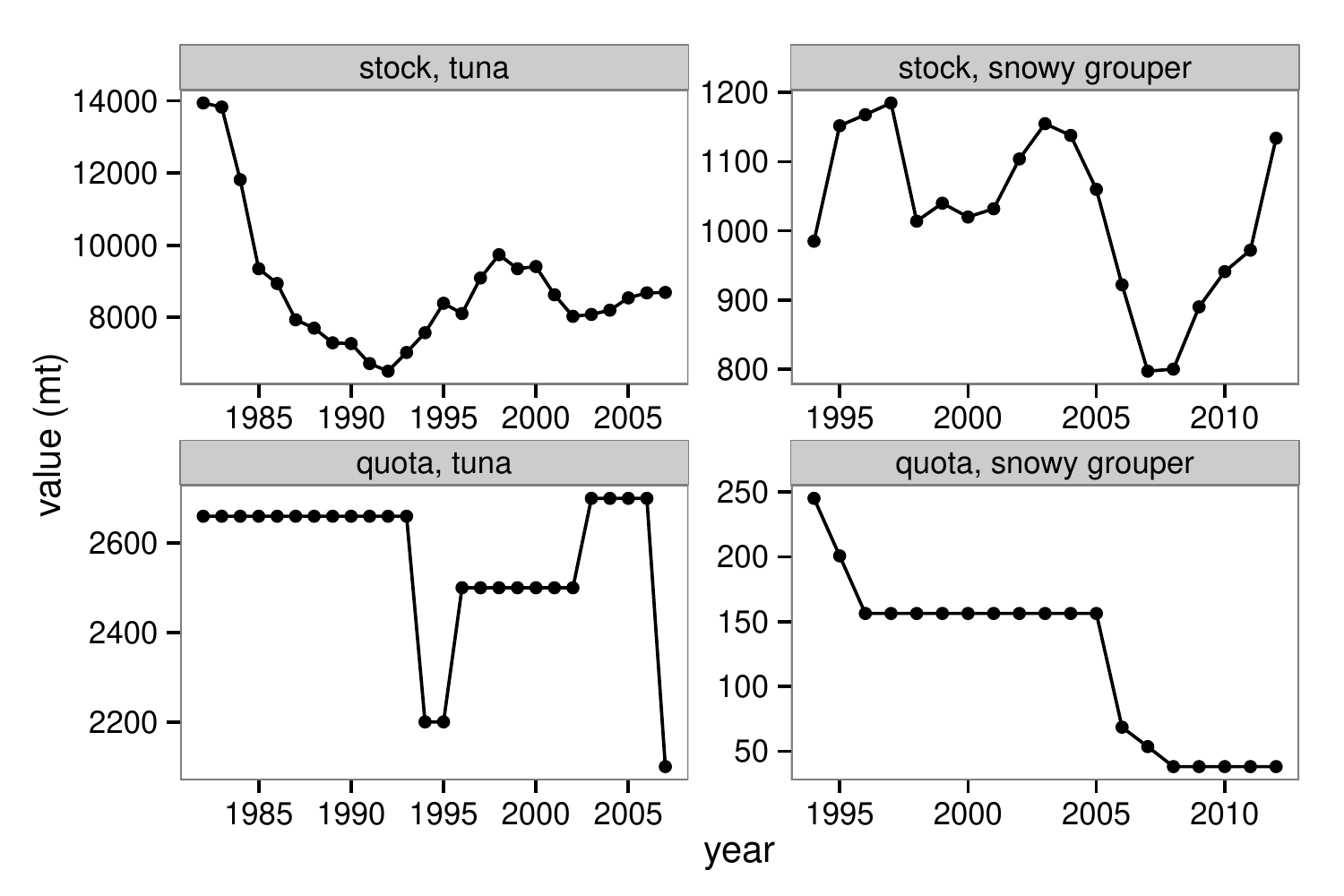}
\caption{Estimated stock abundances (top panels) show much greater
variation than the corresponding harvest policies (bottom). Left panels:
ICCAT Stock assessments and Total Allowable Catch (TAC, before
adjustments for national underages and overages) for bluefin tuna, 1982
- 2007. Right panels: SEDAR stock assessments and annual quotas for
south Atlantic snowy grouper, 1994-2012.}
\end{figure}

To illustrate this tradeoff, we use an example from fisheries
management. Figure 1, top-left panel, shows a time series for the
estimated population size, represented as spawning stock biomass, of
west Atlantic bluefin tuna (\emph{Thunnus thynnus}, henceforth bluefin)
from a recent stock assessment (International Commission for the
Conservation of Atlantic Tunas (ICCAT) 2009). The bottom-left panel
shows the catch quota (Total Allowable Catch or TAC, before being
adjusted for overages or underages in the catch of each country in the
previous year) for the stock as set by the relevant management agency
(International Commission for the Conservation of Atlantic Tunas (ICCAT)
2009). Despite the estimated population size declining by 53.3\% between
1982 and 1991, the quota was not changed during this period. Instead,
the quota only changed occasionally and in between times was left
unaltered. Fishery management decisions regarding this species can be
highly contentious (Safina (1998), Sissenwine et al. (1998), Porch
(2005)). Moreover, the stock is fished by fleets from many nations with
quotas being set by a multilaterial management agency through a process
of negotiation. As such, we might reasonably anticipate that for this
species there could be substantial transaction costs involved in
reaching agreement over any quota change, or constraints on the
frequency with which the those changes could occur. Either could
contribute to the observed quota stability.

Such patterns of variable stocks managed under far less variable quotas
are extremely common, as in the instance of the management of south
Atlantic snowy grouper (\emph{Epinephelus niveatus}) shown in the
righthand panels (SEDAR 2013). Though the stock assessment shows
similarly frequent variation between 1994 and 2012, the quota has only
been adjusted twice during this period. In contrast to the adjustment in
the bluefin quotas, both adjustments in snowy grouper quotas were phased
in through steps over several years: 1994-96 and 2005-08. More
generally, reviews by Biais (1995) and Patterson (2007) document many
cases where changes in catch quotas that a management agency set were
more modest than changes that would be recommended just by considering
variations in stock abundance.

We use a classic fisheries management question to examine how accounting
for costs of policy adjustment can change optimal policies (see also
Ludwig (1980), Feichtinger et al. (1994), Wirl (1999)). We focus on how
harvest quotas for a stochastically varying fish population can be
chosen to maximize the net present value of a fishery. Our formulation
and solution method largely follow Reed's classic treatment on this
question, a treatment repeated widely in bioeconomic textbooks. We note
that while later work has extended this treatment to deal with a variety
of other issues (e.g. Sethi et al. (2005); Singh et al. (2006); McGough
et al. (2009)), we start from the classical model for simplicity of
presentation and analysis.

With his formulation, Reed (1979) showed that a constant escapement
policy could be optimal under certain conditions. Such a policy involves
choosing annual quotas that are perfectly responsive to recruitment
variation in a fish stock. In poor recruitment years, the quota is set
to zero and no fishing is allowed. Any time a recruitment pulse exceeds
the escapement threshold, a quota is set that allows the fishery to
exactly compensate through harvesting, thereby maintaining the optimal
escapement level. However, in that analysis, Reed did not account for
any costs of policy adjustment, which for such a responsive management
strategy potentially could be large.

As the examples in Figure 1 make clear, management policies will rarely
be as responsive as this constant escapement policy assumes. In this
paper, we consider a case where managers seek to balance the benefits in
terms of increased profits from fishing from more finely tracking
recruitment variations with the growing costs associated with adjusting
policies frequently to do so. The policy adjustment costs involved could
reflect pure administrative transaction costs or preferences held by
fishermen, fish processing plants or other stakeholders, for less
variable quotas. In seeking to account for these policy adjustment
costs, we recognize that we do not know just what functional form they
should take and that it will require substantial empirical work to
estimate that. Therefore, we investigate three candidate functional
forms that represent qualitatively different assumptions about how these
costs operate to determine whether the results we obtain are sensitive
to such differences.

\section{Methods}\label{methods}

\subsection{Fish population dynamics (state
equation)}\label{fish-population-dynamics-state-equation}

We will assume Beverton-Holt dynamics with multiplicative environmental
noise

\begin{equation}
N_{t+1} = Z_t \frac{A (N_t - h_t)}{1 + B (N_t - h_t)}, \label{eq:state_equation}
\end{equation}

where \(N_t\) the stock size, \(h_t\) the harvested level, \(Z_t\) gives
the stochastic shocks, which we assume are log-normally distributed and
\(A\) and \(B\) are positive constants.

We assume managers set an annual quota for harvesting \(h_t\) (the
control variable) after observing the stock size that year \(N_t\) (the
state variable), but while still being uncertain about future
environmental conditions and stock sizes. Through time, this gives a
time path of management actions \({\bf h}=(h_1,h_2,\dots)\) that depend
on the stock sizes that were observed (a state dependent control rule).

We assume that managers choose annual quotas to maximize the expected
net present value (NPV) of the fishery. We take as a base case the
situation where there are no costs associated with policy adjustment and
the managers' objective is

\begin{equation}
\max_{{\bf h}}\mathbf{ E} ( NPV_{0} )=\max_{{\bf h}} \sum_0^{T_{\mathrm{max}}} 
\mathbf{E} \left( \displaystyle \frac{\Pi_0(N_t,h_t)}{(1+\delta)^{t+1}} \right) \label{eq:objective}
\end{equation}

where \(\mathbf{E}\) is the expectation operator, \(\delta>0\) is the
discount rate and \(\Pi_0\) is the net revenue from operating the
fishery in a given year. In this base case, we assume the annual
dockside (that is, before accounting for any adjustment costs that may
be incurred) net revenue from fishing is

\begin{equation}
\Pi_0(N_t,h_t) = p h_t -  c_0 E_t \label{eq:annual_netrevenue}
\end{equation}

where \(E_t\) represents fishing effort. We assume catch is proportional
to stock size and effort expended fishing that year, \(h_t = q E_t N_t\)
and constant \(q>0\) is the catchability coefficient. In Eqn.
\eqref{eq:annual_netrevenue}, \(p\) is the price per unit harvest and
\(c_0\) the cost of fishing per unit effort and, to simplify the
presentation of results, we assume that these are constants with \(p>0\)
and \(c_0>0\).

To simplify presentation of the results, we will illustrate cases where
the growth parameters are chosen so that the equilibrium biomass for the
equivalent deterministic model without harvesting is 10; (specifically,
\(A=\) 1.5 and \(B=\) 0.05 in Eqn \eqref{eq:state_equation}). To
characterize environmental variability, we assume multiplicative shock
\(Z_t\) is distributed log-normally with log standard deviation
\(\sigma_g\) = 0.2. In addition, we show cases where \(p=\) 10,
\(\delta=\) 0.05 and \(c_0=\) 30. In the supplement we illustrate that
the qualitative patterns observed here are not sensitive to the specific
choices of these parameter values.

Taken together this objective function \eqref{eq:objective} and the
state equation \eqref{eq:state_equation} define a stochastic dynamic
programming problem that we solve using backwards recursion via
Bellman's equation. We denote the resulting state-dependent, optimal
control as \(\mathbf{h}_0^{\ast}\). We solve this problem on a finite
time horizon of \(T =\) 50 using value iteration (Mangel and Clark 1988,
Clark and Mangel 2000).

R code for implementing the dynamic programming algorithm in this
context is provided as a supplementary R package to the paper,
(\href{https://github.com/cboettig/pdg_control}{github.com/cboettig/pdg\_control},
Boettiger et al. (2015)), along with scripts for replicating the
analyses presented here. To address concerns about computational
reproducibility (Boettiger 2015), a Docker container image of the
software environment is also provided.

\subsection{Costs of policy
adjustment}\label{costs-of-policy-adjustment}

We compare this base case to three alternative problem formulations,
each reflecting different plausible functional forms that costs of
policy adjustment could take. In each, we assume managers can adjust the
quota set in the fishery \(h_t\) in a given year and that any policy
adjustment costs are associated with changes to this control variable.
In each case, we assume there is no cost in initially setting the
harvest policy at time 0.

First we assume that policy adjustment costs are directly proportional
to the magnitude of the change in policy being proposed, such that
larger changes to annual harvesting quotas incur greater policy
adjustment penalties. Specifically, we replace \(\Pi_0\) in the
\(NPV_0\) equation with

\begin{equation}
\Pi_{1}(N_t,h_t, h_{t-1}) = \Pi_0 - c_1  |  h_t - h_{t-1} | \,.
\label{eq:Pi_1}
\end{equation}

Next we continue to assume that policy adjustment costs depend on the
magnitude of the change in policy being proposed. However, we consider a
case where this dependence is nonlinear with big changes in policy being
disproportionately expensive:

\begin{equation}
\Pi_{2}(N_t,h_t, h_{t-1}) = \Pi_0 - c_2 (  h_t - h_{t-1})^2 \,.
\label{eq:Pi_2}
\end{equation}

Finally, we assume instead that there is a fixed cost associated with
any adjustment of policy, regardless of how large that adjustment might
be

\begin{equation}
\Pi_{3}(N_t,h_t, h_{t-1}) = \Pi_0 - c_3 (1-\mathbf{I}(h_t, h_{t-1}))  \,,
\label{eq:Pi_3}
\end{equation}

where indicator function \(\mathbf{I}(h_t, h_{t-1})\) takes the value 1
if \(h_t=h_{t-1}\) and zero otherwise. For each \(\Pi_i\), we can then
define a new objective function \(NPV_i\) similar to that in Eqn.
\eqref{eq:objective}. We note that the fixed cost has conceptual
analogues to the set-up cost of Reed (1974) and Spulber (1982) but here
the fee is associated with a change in harvesting (\(h_t\neq h_{t-1}\))
and not with harvesting \emph{per se} (\(h_t>0\)) as in those studies.

Taken together with the state equation Eqn. \eqref{eq:state_equation},
each of these new objective functions defines a different stochastic
dynamic programming problem. Again, we solve them numerically using
backwards recursion. To include costs of policy adjustment, we expand
the state space to include both the current stock size and the
management action taken on the previous time step \((N_t,h_{t-1})\). For
each new objective function \(\max NPV_i\), we denote the corresponding
optimal control policy as the vector \(\mathbf{h}_i^*\).

Note that this problem is much larger computationally than the classic
formulation of this stochastic dynamic programming problem for a single
stock. Whereas the classic problem considers each of \(Q\) possible
discrete quotas for \(S\) possible stock values at each time \(t\), for
a search space of size \(S \times Q\) per timestep; this formulation
must also consider all possible values of quota in the \emph{previous}
time step: since the costs that will follow depend on whether and by how
much the harvest policy will change. This creates a total of
\(S \times Q \times Q\) configurations.

In the Supplementary Material, we compare our results with more
conventional fisheries economic formulation in which additional costs
are applied to the control variables themselves, as opposed to
adjustments to the controls:
\(\Pi_4 (N_t,h_t) = p h_t - c_0 E_t-c_4E_t^2\). Additional costs of this
form tend to have a smoothing effect on optimal quotas, but also change
the long-term average stock size or quota size that is optimal.

\subsection{Comparing apples to
apples}\label{comparing-apples-to-apples}

Each policy adjustment cost function is characterized in terms of a cost
coefficient \(c_i\). However, \(c_i\) takes different units for each
functional form. Therefore, if seeking to compare the relative effect of
each penalty function on optimal management, it is unclear what
parameter values should be used. To address this issue, we calibrate
choices of \(c_i\) so that each has a comparable impact on the optimal
fishery \(NPV\). Fortunately, the optimal control framework provides a
completely natural way to make this comparison: in terms of the
\emph{economic value} (measured by \(NPV\)) of the stock in each case.
We will calibrate our realized economic value under each policy relative
to economic value of the stock when adjustments to the policy are free
(\(NPV_0\)).

Figure 2 shows the calibration graphically. Each curve plots the change
in maximum expected \(NPV\) for a given cost structure as the \(c_i\)
coefficient is increased. In each case, the figure shows maximum
expected NPV with policy adjustment costs as a proportion of the maximum
expected NPV available in the basic problem \(NPV_0({\bf h_0^*})\)
without policy adjustment costs.

\begin{figure}[htbp]
\centering
\includegraphics{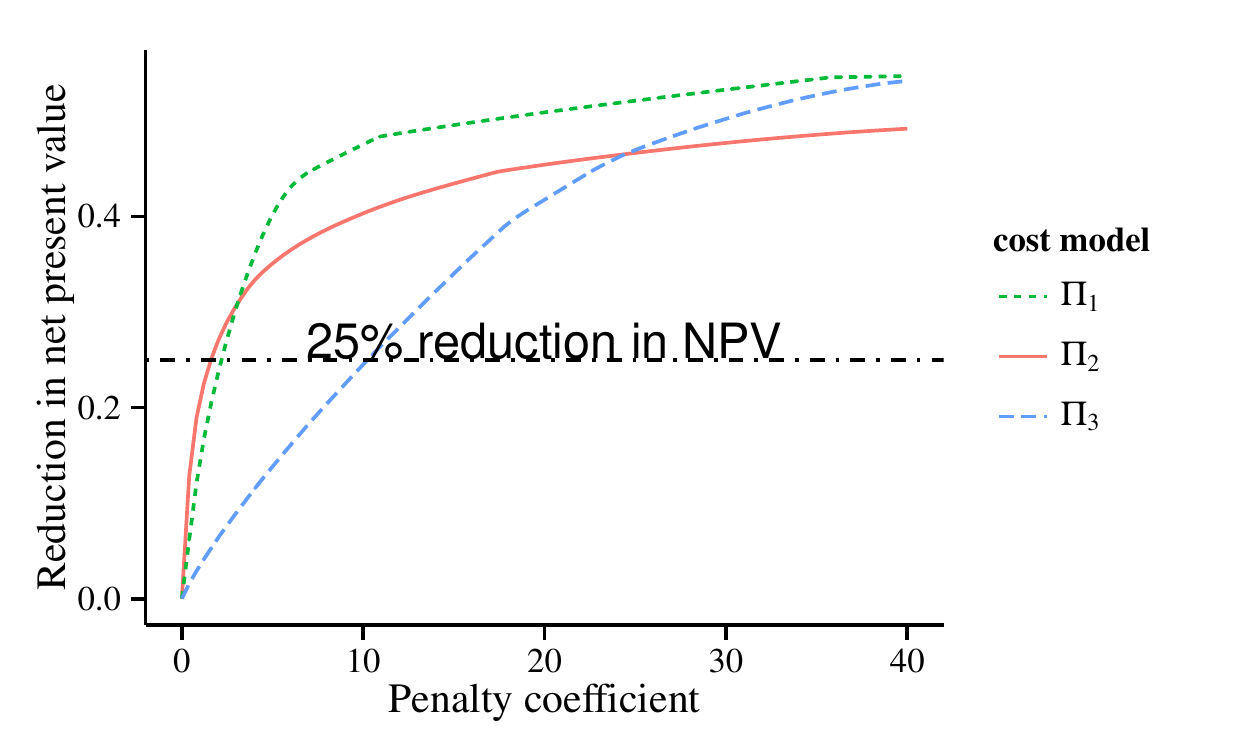}
\caption{Expected net present value by functional form of penalty.
Horizontal axis shows the coefficient \(c_i\) governing the magnitude of
the policy cost, while vertical axis shows fraction of the maximum
expected net present value dissipated by the policy cost,
\(\tfrac{NPV_0 - NPV_i}{NPV_0}\). The horizontal line indicates a value
of the stock that is reduced by 25\% from the maximum expected value in
the absence of policy adjustment costs, \(NPV_0({\bf h_0^*})\).
Selecting the coefficient \(c_i\) corresponding to this value in each
functional form allows us to make consistent comparisons across the
different functional forms of policy costs.}
\end{figure}

To compare the impact of penalty functions on optimal management across
the different functional forms, we select penalty cost coefficients that
induce the same reduction in maximum expected NPV. For example, the
dashed vertical line in Figure 2 maps the needed cost coefficient
\(c_i\), for each penalty function, such that the fishery is worth 75\%
of its unconstrained value when optimally managed. To demonstrate that
the results shown here are independent of this choice of calibrating to
a 75\% reduction, we conduct the same analysis with a 10\% and 30\%
reduction rate as well. The corresponding figures can be found in the
Supplemental Materials.

\section{Results}\label{results}

\subsection{Effect of policy adjustment costs on optimal quotas and
stock
sizes}\label{effect-of-policy-adjustment-costs-on-optimal-quotas-and-stock-sizes}

\begin{figure}[htbp]
\centering
\includegraphics{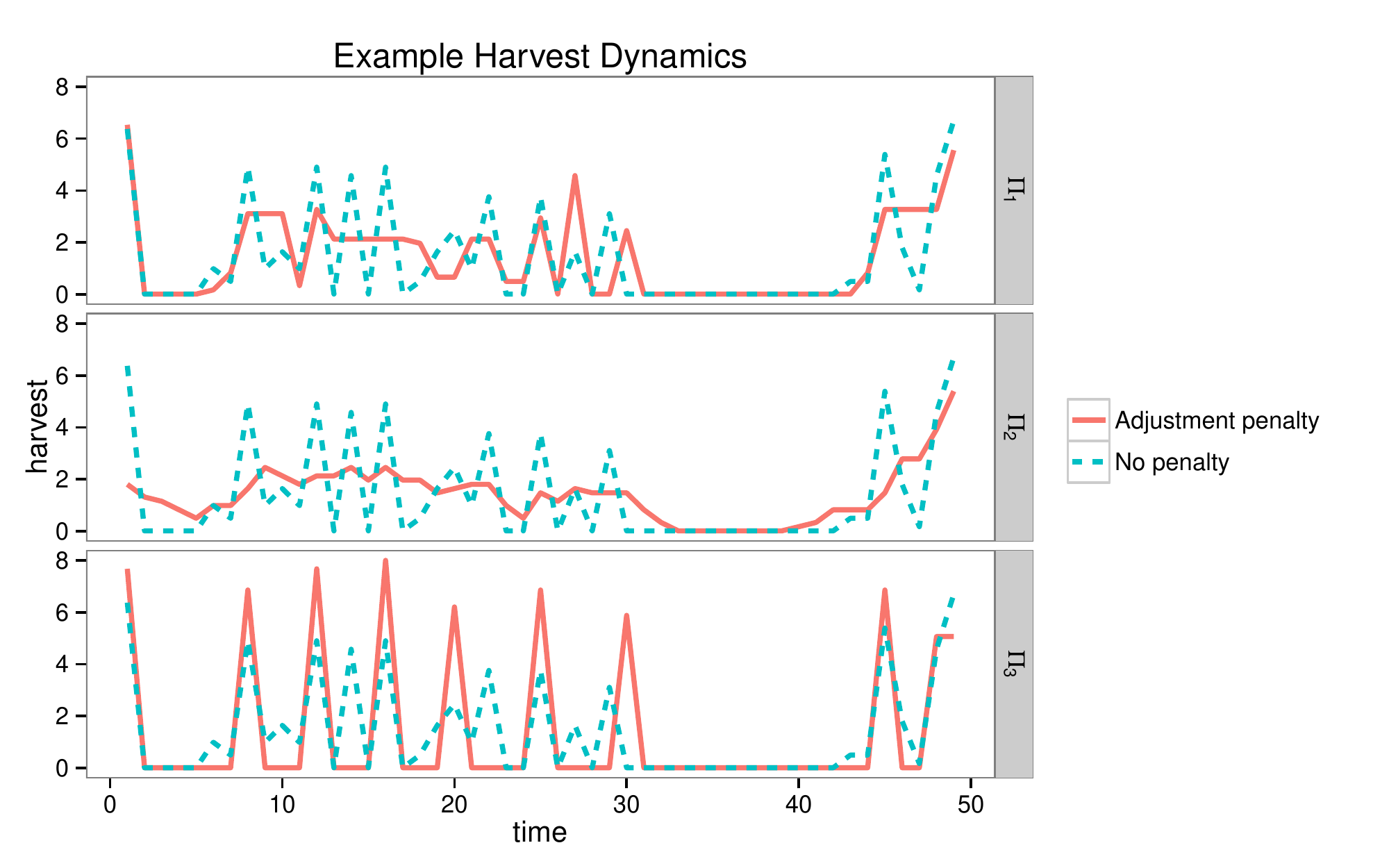}
\caption{Example realization of optimal harvesting strategy under the
different functional forms of adjustment costs. When costs vary linearly
with the size of the adjustment (\(\Pi_1\)), periods with no adjustment
are more common than if adjustments were free. When costs vary
quadratically with adjustment size (\(\Pi_2\)), small adjustments are
relatively cheap and thus the resulting optimal policy always changes,
but by a smaller amount than if adjustments were free. When adjustments
incur a fixed fee independent of size (\(\Pi_3\)), the optimal strategy
either remains unchanged or overshoots the cost-free optimum.}
\end{figure}

Figure 3 illustrates how the different forms of adjustment cost can
impact the dynamics of the optimal harvest policy. Corresponding stock
sizes can be seen in the supplement. Each panel is generated against the
same sequence of environmental variability so that they can be compared
directly. The harvest policy chosen shows a systematic deviations from
the adjustment cost free optimum, depending on the structure of the cost
function.\\In each case, the optimal solution without any adjustment
cost is shown by the dashed grey line, with the policy induced by
optimization under the given cost structure (equivalent to a 25\%
reduction in maximum expected NPV) overlaid in solid blue.

The first panel shows a typical pattern resulting from linear adjustment
costs (\(\Pi_1\)). The optimal policy tends to avoid very small policy
adjustments, resulting in periods of a constant policy followed by
sudden bursts of adjustment. This results in a relatively step-like
policy pattern. In contrast, the second formulation (quadratic costs,
\(\Pi_2\)) disproportionately penalizes large policy adjustments. The
corresponding optimal policy is typified by the middle panel, responding
to each of the fluctuations in stock, made by the cost-free policy, but
with smaller magnitude response than the equivalent cost-free optimal
solution. This results in a smoother \(h_t\) curve, one that undershoots
the larger oscillations seen in the cost-free optimum in favor of a
policy that changes incrementally each year. Finally, the optimal policy
for the third, `fixed fee' formulation (\(\Pi_3\)) only makes large
adjustments, as one might expect, because the magnitude of the
adjustment made is not reflected in the resulting cost.

These patterns are consistent across stochastic replicates over a range
of penalty magnitudes. The comparisons shown in Figure 3 are for one
realization (i.e.~a particular sequence of random number draws
representing environmental variability) and are made for one particular
magnitude of policy adjustment costs, which have been calibrated to
equal to 25\% of \(NPV_0({\bf h_0^*})\). To demonstrate this, we solved
for the optimal policy under each of the three adjustment cost scenarios
for 100 different \(c_i\) coefficients. For each resulting policy, we
then simulated 500 stochastic replicates of the stock dynamics managed
under that policy. To summarize the patterns shown in Figure 3 and
characterize the impact of increasingly large fraction of the economic
value (NPV) being consumed by policy adjustment costs, we examine the
response of several summary statistics to increasing adjustment
penalties (as a fraction of the adjustment-free value) in Figure 4. We
show the impact of policy adjustment costs on the variance \&
autocorrelation of the harvests through time.

\begin{figure}[htbp]
\centering
\includegraphics{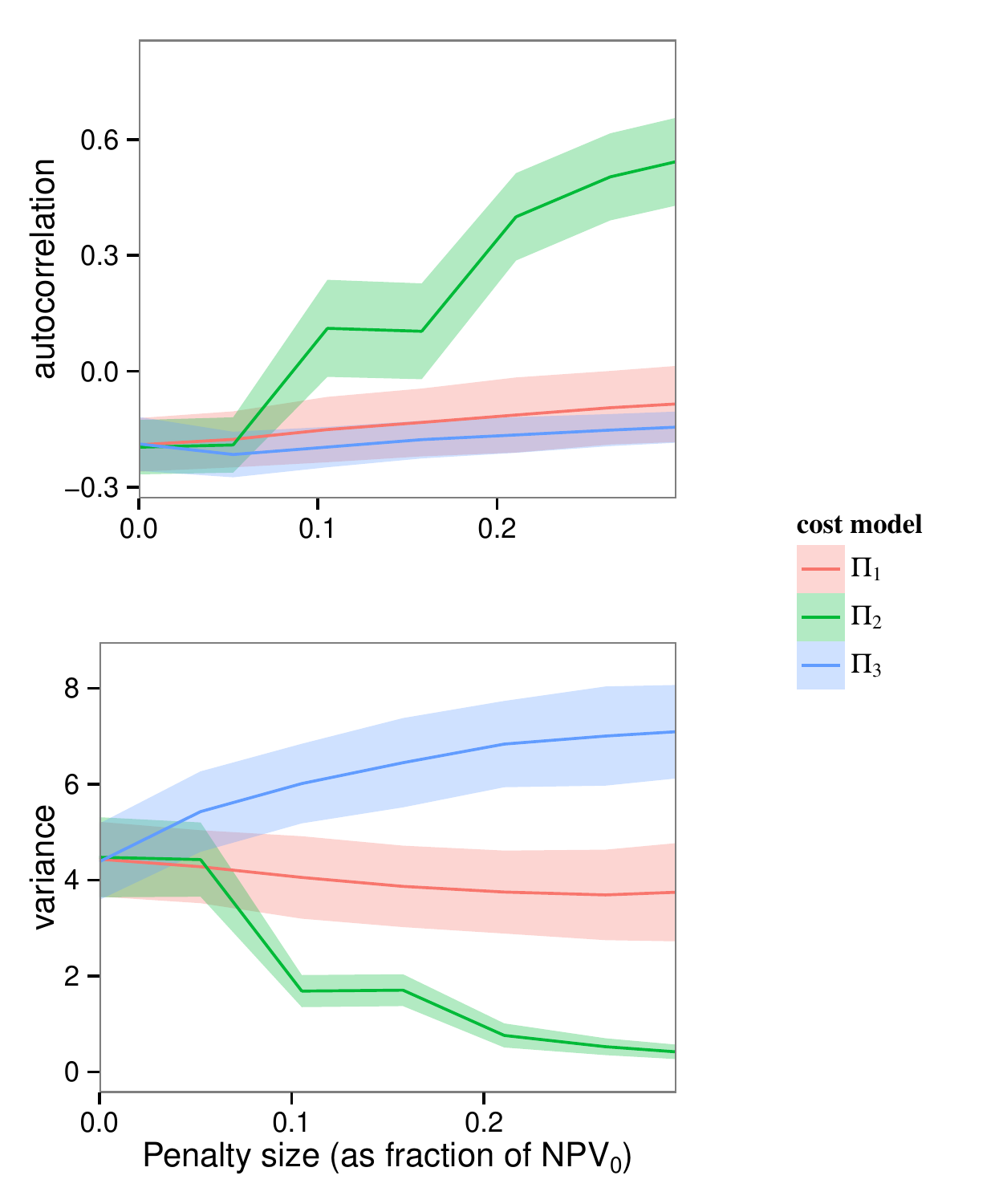}
\caption{Variance and autocorrelation of harvest dynamics, as a function
of the penalty coefficient. Adjustment penalty is calculated as
percentage of the maximum expected NPV in the basic case with no policy
adjustment costs.}
\end{figure}

Despite summarizing across such a large ensemble, Figure 4 confirms the
patterns we observed in the individual realizations shown in Figure 3.
For example, when small policy adjustments cost little but large
adjustments are expensive (\(\Pi_2\)), we see the smoothing signal that
we might have expected (see also Ludwig (1980) for this particular
case). As cost penalties increase in severity (moving right along the
horizontal axis, larger fractions of the economic value are consumed by
adjustment costs), the variance in quotas through time decreases and the
autocorrelation of quotas through time increases. When adjustment costs
are zero, all policies are slightly negatively autocorrelated: this
reflects the fact that a high harvest year is usually followed by a low
harvest year that allows the stocks to recover. However, as adjustment
costs increase, the quadratic policy tends to undershoot when stocks
jump to very high or low levels, adjusting to compensate over multiple
years, as seen in Figure 3. As a result, autocorrelation increases.

Interestingly, as suggested by the realization in Fig. 3, including a
fixed cost (\(\Pi_1\)) of policy adjustment increases the variation in
harvests through time. This is the opposite of a smoothing effect.

Finally, the case where policy adjustment costs scale linearly with the
size of the adjustment (\(\Pi_1\)) appear to be something of a middle of
the road strategy, in that increasing the severity of policy adjustment
costs. To reveal the particular impact of policy adjustment costs of
this type requires a more targeted summary statistic. Specifically, for
each run we calculated the frequency with which the optimal policy
involved maintaining a positive quota across multiple time steps
unaltered. This type of policy is arguably the most commonly observed
behavior in TAC management, but is one that is very rarely observed to
be part of the optimal management strategy in the basic model without
policy adjustment costs (Eqn. 3) or when optimizing against \(\Pi_2\) or
\(\Pi_3\).

Figure 4 shows these qualitative patterns still hold at when the models
are calibrated to a range of different reductions in NPV, from 0\% to
30\%, though naturally the magnitude of the differences is greatest when
the penalty is larger. Note that the supplementary material provide
further evidence that these patterns are independent of the exact
reduction shown by providing results analogous to Figure 3 (S8, S9) and
Figure 5 (S10, S11), and Figure 6 (S6).

While still not common for the particular parameter combinations we
examine, we find that positive unaltered quotas through time are much
more likely to occur when optimizing against \(\Pi_1\) where policy
adjustment costs scale linearly with the size of proposed policy
changes: over the 100 replicates time series the harvest policy is
strictly positive and identical in consecutive intervals only 9.88\% of
the time for quadratic costs \(\Pi_2\), compared with 20.78\% with
linear costs \(\Pi_1\) and 0.37\% for fixed costs \(\Pi_3\). Moreover,
these occurrences increase in frequency as the severity of these costs
(\(c_i\)) increases. Note that the restriction of only positive quotas
lets us distinguish between cases that are constant due purely to
adjustment costs from cases that are constant purely due to boundary
effects.

\subsection{Consequences of policy adjustment
costs}\label{consequences-of-policy-adjustment-costs}

\begin{figure}[htbp]
\centering
\includegraphics{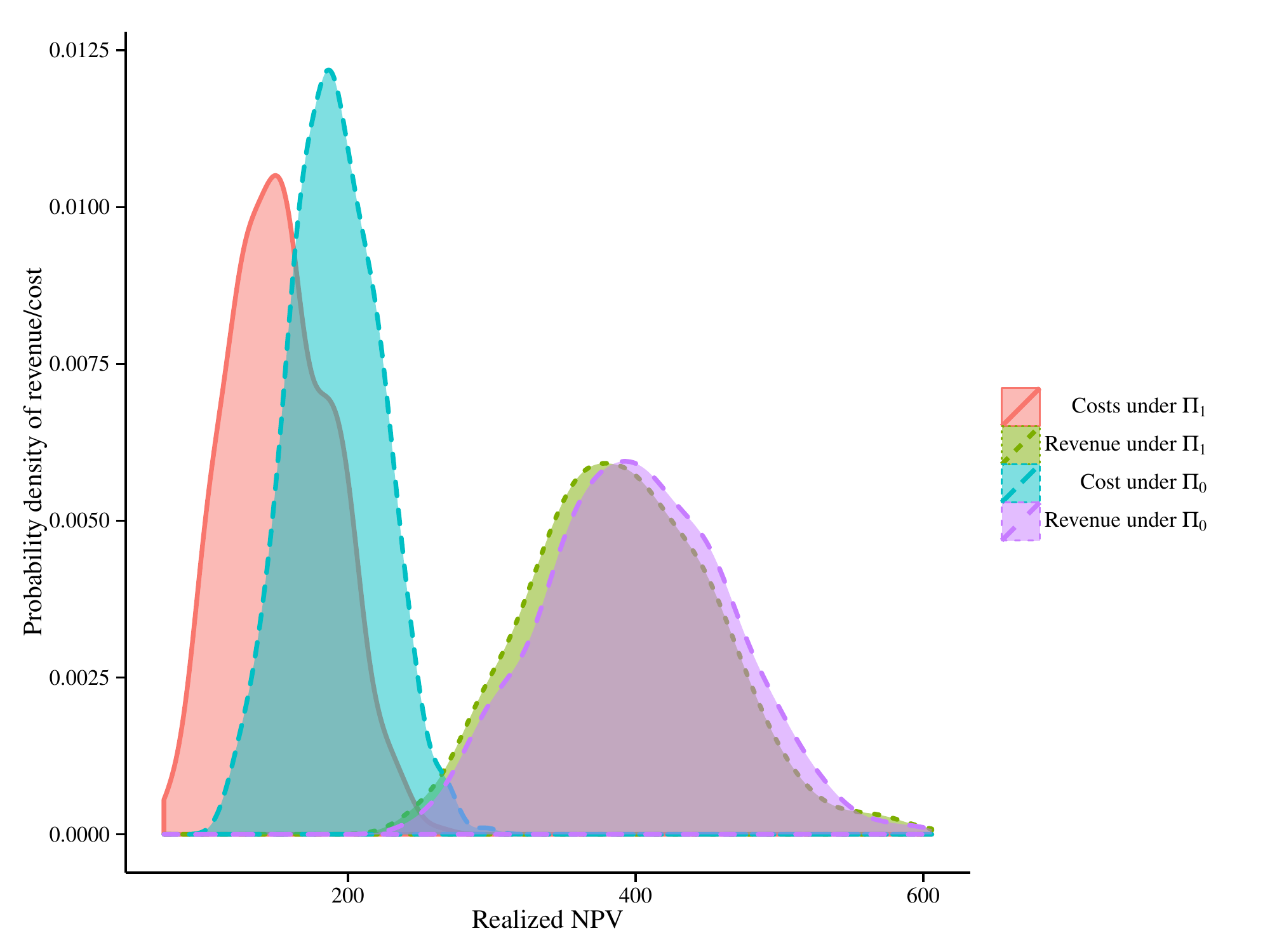}
\caption{Distribution of revenues from fishing and costs paid to
adjustment. Plot shows distributions for the linear costs structure
(\(\Pi_1\)) only, alternate penalties can be found in the supplementary
materials. Dockside revenues are always higher than adjustment costs.
When those costs are not accounted for in the policy (`no adjustment
costs'), it is possible to obtain only marginally higher revenues, but
pay higher adjustment costs.}
\end{figure}

Next we examine the consequences either of ignoring policy adjustment
costs when they are present or assuming they are present when they are
not.

To do so, we will simulate managing a fish stock when adjustments are
costly. We will perform 500 replicate simulations of this for each of
the three cost structures, \(\Pi_1\), \(\Pi_2\), \& \(\Pi_3\). For each
replicate, we will use two different management policies: one which
ignores the adjustment costs, (that is, the Reed optimum policy,
\(\Pi_0\)), and a second which accounts for the adjustment cost in
place.

By comparing the NPV of this second scenario against the first, we can
quantify the impact of ignoring costs when they are present.

Figure 5 shows that the difference in dockside revenue (labeled
``profits'' in the figure) between these two scenarios (green vs purple
distributions) is very small, while a larger difference exists between
the costs paid for making policy adjustments (red vs blue). This is
predicted by the optimization: the first scenario must outperform (or at
least equal) the NPV of the second scenario in aggregate, and as it
cannot do so through higher profits (these having already been maximized
by the cost-free strategy), it must do so through lower costs.
Interestingly, for this set of parameters very little is lost in
revenue. This suggests that the impact of assuming costs are present
when they are not is rather small, as the revenue is nonetheless near
optimal. This also suggests that the converse error -- ignoring costs
when they are present -- is more severe.

\begin{figure}[htbp]
\centering
\includegraphics{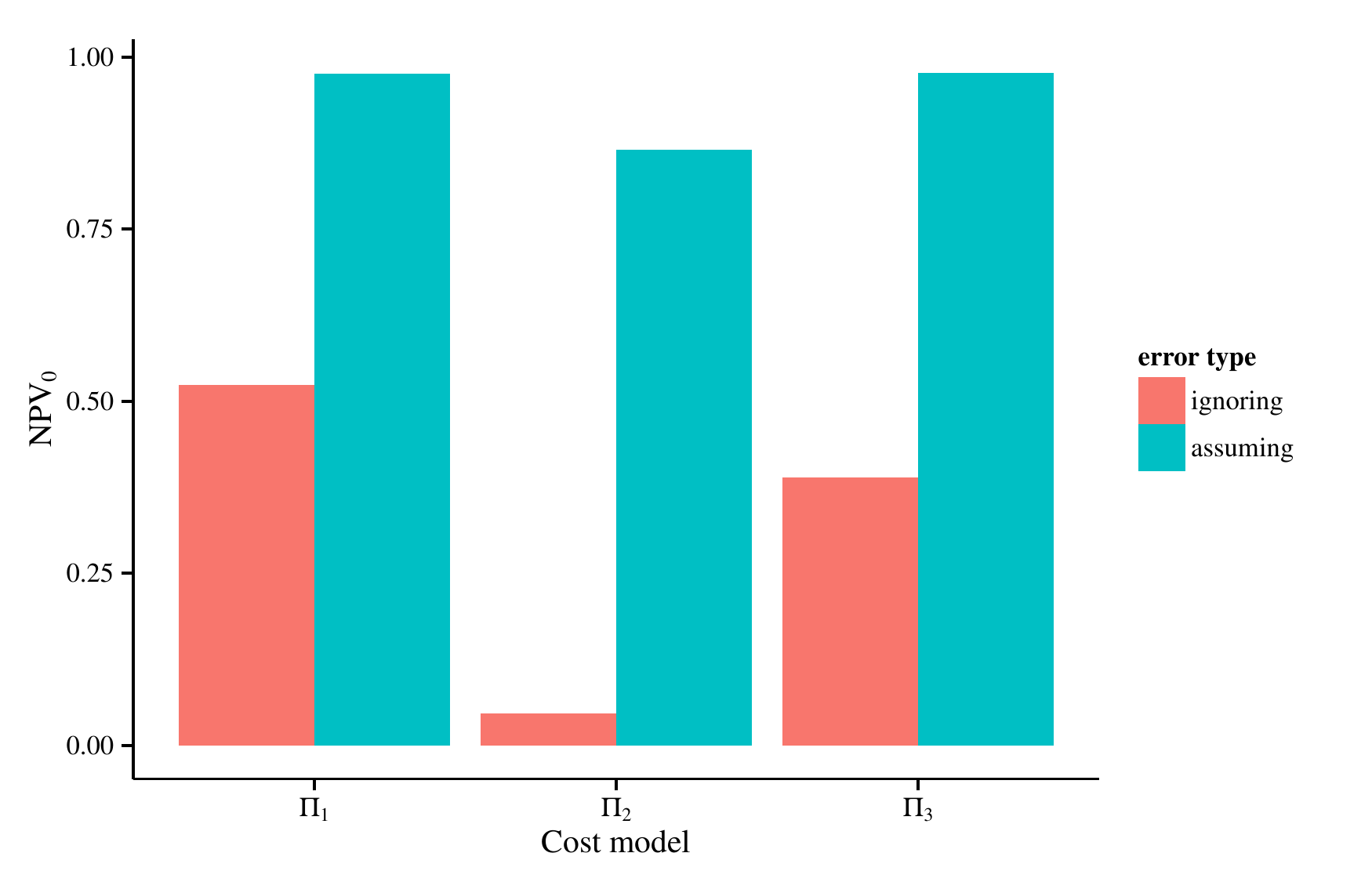}
\caption{Relative cost of ignoring adjustment costs when they are
present (`ignoring', red) vs assuming the adjustment costs when they are
absent (`assuming', blue). All values are relative to the cost free
adjustment optimum, \(NPV_0\). In each case, it is less costly to assume
adjustment costs are present when they are not than it is to ignore them
when they are present.Supplementary Figure S6 shows how the magnitude of
this effect changes with the size of the reduction to \(NPV\).}
\end{figure}

We illustrate the differences between these two errors directly in
Figure 6. Subtracting revenue from adjustment costs and considering only
the mean across the replicates, we can derive the expected \(NPV\) in
each case. To standardize against a common baseline, we will express
this value as a fraction of the cost free optimum, \(NPV_0\). Note that
we have calibrated the optimal solutions to anticipate that 25\% of
their \(NPV\) would go into adjustment costs. When these adjustment
costs are not actually present (blue bars, Fig 6), we thus expect a
\(NPV\) higher than the anticipated 75\% but not quite as high as the
optimal policy for that scenario (100\%). When adjustment costs are
present but the policy assumes that they are not (red bars), we would
expect the \(NPV\) to be less than 75\% of \(NPV_0\) (since that's as
good as the optimal solution for this scenario can do).

Figure 6 shows that across each penalty form \(\Pi_1\), \(\Pi_2\) \&
\(\Pi_3\), the economic impact of ignoring adjustment costs when they
are present is higher than accounting for them when they are absent.
This holds even after accounting for effect described above: the red
bars are consistently well below the 75\% of \(NPV_0\), while the blue
bars are barely below \(NPV_0\) (a value of 1 on the vertical axis). We
also see that the impacts on net present value of the fishery of
incorrectly assuming or ignoring policy adjustment costs are most severe
when these costs are assumed to scale quadratically with the size of the
policy change being implemented (\(\Pi_2\)). In the Supplementary
Materials we show that pattern holds across different severity of
adjustment costs (robust to choice of \(c_i\)), though the larger the
adjustment cost, the greater the difference between these two errors.

Note that this analysis does not consider the possibility that an
adjustment cost is present, but of a different functional form than the
policy assumes. The consequences of managing under the wrong functional
form can be as or more severe than simply ignoring adjustment costs in
the optimization altogether. We illustrate several of these mismatch
errors in the Supplementary Material.

\section{Discussion}\label{discussion}

Policymakers managing ecological systems that vary in space and time
must evaluate how much of that variation to reflect in management
recommendations. Finer tuning a policy to respond to frequent variations
in ecological dynamics may incur increased transaction costs associated
with constantly revisiting past policy decisions. A more pragmatic
approach would be one that balances benefits from responding to frequent
variations in ecological conditions with the increased transaction costs
involved. As a first step towards exploring these ideas, we revisited a
classic problem from bioeconomics concerning the optimal management of a
fish stock subject to stochastically varying recruitment (Reed 1979,
Clark 2010). We examined how optimal policy recommendations -- here
annual catch quotas -- changed when accounting for costs associated with
policy adjustment and what the implications of following these policy
recommendations would be for the exploited population. We also compared
how the value of the fishery is affected by managers either under- or
overestimating the importance of policy adjustment costs of this type.

Estimating policy adjustment costs and how they respond to the size of
proposed policy changes would be empirically challenging. Recognizing
this fact, we compared different plausible forms that these costs might
take. We compared two cost structures where we assumed the magnitude of
policy adjustment costs increased with the magnitude of the change in
policy being proposed with an alternative formulation in which we
assumed there was a fixed cost associated with making any change to
current policy. While each formulation provides only very
phenomenological representations of the policy-setting environment, we
believe that each enables us to explore meaningful differences in how
policy adjustment costs might operate. That being said, in the real
world we might expect the different types of adjustment cost to operate
in combination.

The biggest differences between the representations of policy adjustment
cost that we consider are between those that have a smoothing effect on
annual quotas and those that do not. The few past studies that
incorporate policy adjustment costs in models of fisheries or other
environmental management contexts (Feichtinger et al. 1994, Wirl 1999)
have assumed costs of policy adjustment increased as a quadratic
function of the magnitude of the policy change being proposed (analogous
to \(\Pi_2\) formulation). In effect, this means that small changes to
annual quotas incur little extra cost, but large changes to annual
quotas become disproportionately expensive to make. Including costs of
this form smooths inter-annual variation in the recommended catch
quotas. Also, the catches and remaining stock sizes corresponding to
optimal management become more autocorrelated in time, because it takes
the fishery multiple timesteps to harvest down peaks in abundance that
follow large recruitment pulses. The effects of smoothing here are
similar to those predicted when assuming the cost per unit effort is
increasing in the amount of effort expended (Brown 1974, Lewis 1981,
McGough et al. 2009); (see Supporting Information for a summary of the
relevant results) as opposed to associating extra costs with changes to
policy \emph{per se}. Smoothing effects of this type are what one might
have expected when including policy adjustment costs.

Standing in sharp contrast to these smoothing predictions is our finding
that including policy adjustment costs can actually \emph{increase} the
variability of quotas through time if there is a fixed cost associated
with making any changes to current policy (e.g.~costs of running
relevant stakeholder meetings and public consultations on proposed
policy changes, \(\Pi_3\)). With this formulation, as stock sizes vary
in response to recruitment, the fishery manager must balance the cost of
brokering a policy change with the cost in forgone fisheries revenue
from not responding to favorable recruitment pulses. The optimal policy
involves ignoring small variations in recruitment, but then assigns a
larger quota when particularly strong recruitment years arise than would
have been the case in the absence of policy adjustment costs. The other
functional form we consider, in which the costs of policy adjustment
scale linearly with the size of the adjustment (\(\Pi_1\)), has less
obvious effects on optimal policies. Interestingly, it is this structure
that most frequently produces stretches of strictly positive but
unchanging quotas of the type most commonly encountered in real world
applications.

That the different representations of policy adjustment costs result in
such different dynamics suggests that researchers constructing fisheries
economics models should proceed cautiously when choosing how to
represent these costs. However, in our own experience, we have found
that while modeling studies sometimes mention policy adjustment costs
when motivating model assumptions (Brown 1974, Lewis 1981, Feichtinger
et al. 1994, Wirl 1999, McGough et al. 2009), they rarely provide much
justification for the choice of functional form used or test the
sensitivity of any conclusions drawn to alternative specifications.
Indeed, we were surprised to find such clear differences between the
functional forms, because we anticipated that the limiting case of
increasing policy adjustment costs within each functional form should be
the same, namely a constant annual quota that does not change through
time.

We also compared the efficiency costs that would result from failing to
account for policy adjustment costs, if they are present, with those
involved in assuming them when in fact they are absent. The results of
this comparison were not sensitive to the particular form of policy
adjustment costs assumed. Instead, we always found the efficiency costs
of ignoring policy adjustment costs when they were present to be much
larger than the efficiency costs of assuming policy adjustment costs
applied when they were in fact absent. Policy adjustment costs affect
the overall value of the fishery in two ways here. First, there is the
direct cost associated with each quota change. Second there is a cost in
foregone revenue from missed catches when not following what would be
the optimal policy if quotas were free to track recruitment variability.
Our finding that it is more costly to ignore policy adjustment costs
when they are present arises because the first, more direct, cost
contribution here is the larger.

Assuming an adjustment cost exists when in fact it is absent is not the
same as using the wrong functional form when it is present. In the
supplement we show that choice of the functional form makes a
significant difference. We show it can be better to ignore policy costs
than to derive a policy based on assuming the wrong functional form.
This result further underscores the importance of modelers using caution
when seeking to account for these costs. Assuming an arbitrary form in
order to capture the influence of adjustment costs is thus unlikely to
be instructive.

As with any modeling analysis, our formulation makes many assumptions.
For example, we assume that the fishery in question is being optimally
managed by a policymaker who acts as the ``sole owner'' of the stock.
This approach is different to models that assume fisheries are not well
managed, e.g.~by assuming regulated open access conditions (Homans and
Wilen 1997), or models that derive policy recommendations endogenously
by modeling strategic interactions between different stakeholders
(Kaitala 1993, Laukkanen 2003). As such, we anticipate that our approach
will be more relevant to some fisheries, particularly domestic fisheries
in developed countries that are subject to relatively strong regulatory
regimes, than to others (e.g., artisanal fisheries that are subject to
weaker regulation). To focus on the effects of policy adjustment costs,
we focused narrowly on the basic model specification of Reed (1979) and
examined how the predictions of this classic problem were changed by
introducing policy adjustment costs of different forms. However, there
have been many elaborations on Reed's basic approach that increase the
realism of the optimization models involved by relaxing other
assumptions (see for example Sethi et al. (2005); Singh et al. (2006);
McGough et al. (2009)).

One obvious research avenue suggested by our models is that of
empirically estimating costs of policy adjustment. A direct estimation
approach could quantify some sources of policy adjustment costs,
e.g.~costs to processing plants that arise from having more variable
catches. However, other more intangible sources of policy adjustment
costs, e.g.~preferences of policymakers or different stakeholders for
less variable quotas, might be missed. An alternative, more holistic,
approach would be to apply revealed preference methods to fisheries
management agencies themselves, in the tradition of (McFadden 1975,
1976). Such an analysis would involve comparing quotas that were set
relative to stock sizes as they were estimated at the time each
management decision was taken to try to infer what objective managers
were maximizing.

A worthwhile modeling extension suggested by our results would be to
examine the implications of policy adjustment costs for risks of stock
collapse. This includes models with larger stochasticity as well as
model formulations that can permit alternative stable states. Examining
risks of stock collapse is not possible with the Reed model formulation
that we followed here (both cases are inconsistent with the assumptions
of Reed (1974)) but would be worthwhile in light of our finding that the
variance and autocorrelation in stock sizes through time can be affected
when accounting for costs of policy adjustment. Both are properties
known in other modeling settings to be associated with changes to the
risk of extinction or of transitions between alternative stable states
(Scheffer et al. 2009). This highlights the risk that changes in
management policies can either mimic or mask such possible early warning
signs of sudden transitions, and points to another way in which more
context is needed before such approaches can be useful to management
(Boettiger et al. 2013).

When facing highly variable ecological systems, how often should natural
resource managers respond? A highly interventionist strategy would track
ecosystem variation very closely. Alternatively, a manager might choose
only to take action or change policy only when conditions look very
different to those previously experienced. We took a modeling approach
to begin to explore these ideas. Specifically, we focused on a
well-known problem from fisheries management and examined how optimal
management recommendations changed when we accounted for costs
associated with frequently changing management decisions. While we
focused on a fisheries context, our findings would be relevant to many
other settings where natural resource managers revisit management
decisions through time in light of ecological variability, including
game management, management of instream flow rates, fire management,
habitat restoration, and managing for endangered species. The analyses
that we present can also be thought of as providing a temporal
counterpart to discussions about the spatial scale over which ecosystem
management.

\section{Acknowledgements}\label{acknowledgements}

This work was assisted through participation in ``Pretty Darn Good
Control'' Investigative Workshop at the National Institute for
Mathematical and Biological Synthesis, sponsored by the National Science
Foundation through NSF Award \#DBI-1300426, with additional support from
The University of Tennessee, Knoxville. The authors acknowledge helpful
discussions and input from working group co-organizers Megan Donahue and
Carl Toews, and participants Marie-Josee Fortin, Dan Ryan, Frank Doyle,
Claire Paris, Iadine Chades and Mandy Karnauskas, as well as the two
anonymous reviewers whose feedback strengthened this paper. We also
acknowledge the support of NSF Grant DBI-1306697 to CB and ARC DECRA
(DE130100572) \& the ARC Centre of Excellence for Environmental
Decisions to MB.

\section*{References}\label{references}
\addcontentsline{toc}{section}{References}

Armsworth, P. R., and J. E. Roughgarden. 2003. The economic value of
ecological stability. Proceedings of the National Academy of Sciences of
the United States of America 100:7147--51.

Armsworth, P. R., B. A. Block, J. Eagle, and J. E. Roughgarden. 2010.
The role of discounting and dynamics in determining the economic
efficiency of time-area closures for managing fishery bycatch.
Theoretical Ecology.

Biais, G. 1995. An evaluation of the policy of fishery resources
management by TACs in European Community waters from 1983 to 1992.
Aquatic Living Resources 8:241--251.

Boettiger, C. 2015. An introduction to Docker for reproducible research,
with examples from the R environment. ACM SIGOPS Operating Systems
Review 49:71--79.

Boettiger, C., M. Bode, J. N. Sanchirico, J. LaRiviere, A. Hastings, and
P. Armsworth. 2015, July. Repository for: Optimal management of a
stochastically varying population when policy adjustment is costly.

Boettiger, C., N. Ross, and A. Hastings. 2013. Early warning signals:
the charted and uncharted territories. Theoretical Ecology.

Brown, G. 1974. An optimal program for managing common property
resources with congestion externalities. Journal of Political Economy.

Clark, C. W. 2010. Mathematical Bioeconomics: The Mathematics of
Conservation. Third editions. Wiley; Sons, Hoboken, NJ, USA.

Clark, C. W., and M. Mangel. 2000. Dynamic state variable models in
ecology. Oxford University Press, Oxford.

Durrett, R., and S. A. Levin. 1994. The importance of being discrete
(and spatial). Theoretical Population Biology 46:363--394.

Feichtinger, G., A. Novak, and F. Wirl. 1994. Limit cycles in
intertemporal adjustment models. Journal of Economic Dynamics and
Control 18:353--380.

Halpern, B. S., C. White, S. E. Lester, C. Costello, and S. D. Gaines.
2011. Using portfolio theory to assess tradeoffs between return from
natural capital and social equity across space. Biological Conservation
144:1499--1507.

Homans, F. R., and J. Wilen. 1997. A model of regulated open access
resource use. JEEM 32:1--21.

International Commission for the Conservation of Atlantic Tunas (ICCAT).
2009. Report of the 2008 ICCAT bluefin stock assessment session. ICCAT
Collective Volume of Scientific Papers 64:1--352.

Kaitala, V. 1993. Equilibria in a stochastic resource management game
under imperfect information. European Journal of Operational Research
71:439--453.

Laukkanen, M. 2003. Cooperative and non-cooperative harvesting in a
stochastic sequential fishery. Journal of Environmental Economics and
Management 45:454--473.

Lewis, T. R. 1981. Exploitation of a renewable resource under
uncertainty. The Canadian Journal of Economics 14:422.

Ludwig, D. 1980. Harvesting strategies for a randomly fluctuating
population. Journal du Conseil 39:168--174.

Mangel, M., and C. W. Clark. 1988. Dynamic modeling in behavioral
ecology. (J. Krebs and T. Clutton-Brock, Eds.). Princeton University
Press, Princeton.

McFadden, D. 1975. The revealed preferences of a government bureaucracy:
Theory. The Bell Journal of Economics 6:401.

McFadden, D. 1976. The revealed preferences of a government bureaucracy:
Empirical evidence. The Bell Journal of Economics 7:55.

McGough, B., P. A. J., and C. Costello. 2009. Optimally managing a
stochastic renewable resource under general economic conditions. The
B.E. Journal of Economic Analysis \& Policy 9:1--31.

Neubert, M. G. 2003. Marine reserves and optimal harvesting. Ecology
Letters 6:843--849.

Patterson, K. 2007. Social resistance to the obvious good: A review of
responses to a proposal for regulation of european fisheries. American
Fisheries Society Symposium:587--596.

Patterson, K., and M. Resimont. 2007. Change and stability in landings:
the responses of fisheries to scientific advice and TACs. ICES Journal
of Marine Science 64:714--717.

Porch, C. E. 2005. The sustainability of western Atlantic bluefin tuna:
a warm-blooded fish in a hot-blooded fishery. Bulletin of Marine Science
76:363--384.

Reed, W. J. 1974. A stochastic model for the economic management of a
renewable animal resource. Mathematical Biosciences 22:313--337.

Reed, W. J. 1979. Optimal escapement levels in stochastic and
deterministic harvesting models. Journal of Environmental Economics and
Management 6:350--363.

Safina, C. 1998. Song for the blue ocean: Encounters along the world's
coasts. Henry Holt, New York.

Sanchirico, J. N., M. D. Smith, and D. W. Lipton. 2008. An empirical
approach to ecosystem-based fishery management. Ecological Economics
64:586--596.

Scheffer, M., J. Bascompte, W. A. Brock, V. Brovkin, S. R. Carpenter, V.
Dakos, H. Held, E. H. van Nes, M. Rietkerk, and G. Sugihara. 2009.
Early-warning signals for critical transitions. Nature 461:53--9.

SEDAR. 2013. SEDAR 36 stock assessment report: south Atlantic snowy
grouper. Page 146. SEDAR, North Charleston SC.

Sethi, G., C. Costello, A. Fisher, M. Hanemann, and L. Karp. 2005.
Fishery management under multiple uncertainty. Journal of Environmental
Economics and Management 50:300--318.

Singh, R., Q. Weninger, and M. Doyle. 2006. Fisheries management with
stock growth uncertainty and costly capital adjustment. Journal of
Environmental Economics and Management 52:582--599.

Sissenwine, M. P., P. M. Mace, J. E. Powers, and G. P. Scott. 1998. A
Commentary on western Atlantic bluefin tuna assessments. Transactions of
the American Fisheries Society 127:838--855.

Spulber, D. F. 1982. Adaptive harvesting of a renewable resource and
stable equilibrium. Pages 117--139 \emph{in} L. Mirman and D. Spulber,
editors. Essays in the economics of renewable resources. North-Holland
Publishing, Amsterdam.

Walters, C. J. 1978. Some dynamic programming applications in fisheries
management. \emph{in} J. Petkau, editor. Dynamic programming and its
applications. Academic Press.

Wirl, F. 1999. Complex, dynamic environmental policies. Resource and
Energy Economics 21:19--41.
\end{flushleft}

\end{document}